\documentclass[review]{elsarticle}

\usepackage{lineno,hyperref}
\modulolinenumbers[5]

\journal{Electric Power Systems Research}

\usepackage{latexsym}

\usepackage[cmex10]{amsmath}

\usepackage{algorithmic}

\usepackage{array}

\usepackage{bm}   
\usepackage{multirow} 
\usepackage{bigdelim}
 
\usepackage{booktabs}

  \usepackage[caption=false,font=normalsize,labelfont=sf,textfont=sf]{subfig}

\usepackage{stfloats}

\begin{document}

\begin{frontmatter}

\title{Impact of Power System Partitioning on the Efficiency of Distributed Multi-Step Optimization}

\author[mymainaddress]{Junyao Guo \corref{mycorrespondingauthor}}
\cortext[mycorrespondingauthor]{Corresponding author}
\ead[url]{junyaog@andrew.cmu.edu}

\author[mysecondaryaddress]{Gabriela Hug}
\ead[url]{ghug@ethz.ch}

\author[mymainaddress]{Ozan Tonguz }
\ead[url]{tonguz@ece.cmu.edu}

\address[mymainaddress]{Department of Electrical and Computer Engineering, Carnegie Mellon University, Pittsburgh, PA, USA}
\address[mysecondaryaddress]{Power Systems Laboratory, ETH Z{\"u}rich, Z{\"u}rich, Switzerland }

\title{Impact of Power System Partitioning on the Efficiency of Distributed Multi-Step Optimization}

\begin{abstract}
Recent studies have shown that multi-step optimization based on Model Predictive Control (MPC) can effectively coordinate the increasing number of distributed renewable energy and storage resources in the power system. However, the computation complexity of MPC is usually high which limits its use in practical implementation. To improve the efficiency of MPC, in this paper, we apply a distributed optimization method to MPC. The approach consists of a partitioning technique based on spectral clustering that determines the best system partition and an improved Optimality Condition Decomposition method that solves the optimization problem in a distributed manner. Results of simulations conducted on the IEEE 14-bus and 118-bus systems show that the distributed MPC problem can be solved significantly faster by using a good partition of the system and this partition is applicable to multiple time steps without frequent changes.
\end{abstract}


\end{frontmatter}

\linenumbers
{\renewcommand\baselinestretch{1.42}\selectfont
\section{Introduction}
\label{introduction}
With an increasing number of intermittent energy resources and storage devices integrated into the power system, the question that arises is how to optimally coordinate these resources to overcome the uncertainty introduced by the intermittent resources and the inter-temporal coupling of storages. Approaches based on Model Predictive Control (MPC) are able to address these challenges effectively as they determine the current optimal states of the controllable devices with a look-ahead scheme that accounts for the temporal characteristics of both intermittent resources and storage devices. For example, the energy dispatch problem with intermittent resources is formulated as a multi-step optimization problem over a pre-defined time horizon which treats the intermittent energy resources as negative loads that must be consumed when available \cite{xie2008model}. Similar formulation has also been proposed for the AC Optimal Power Flow (AC OPF) problem that aims to minimize the generation cost of non-intermittent generations over a finite time horizon with the integration of wind generation and storages \cite{baker2012optimal}. Both studies have shown that the total generation cost can be reduced by using such an MPC based multi-step optimization approach to effectively coordinate the intermittent resources and storages.

However, the MPC approach is computationally expensive since the size of the optimization problem grows drastically as the optimization horizon increases. Such computation complexity restricts the practical use of MPC because no control actions can be taken if an MPC problem is not solved within the required amount of time due to the lack of computation capability or storage capacity at the central computation entity. To address this issue, distributed MPC has been studied and applied to various applications such as optimal power flow \cite{Kyri2015MPC}, dispatch of generation with emission limitation \cite{elaiw2012application} and automatic generation control \cite{venkat2008distributed}. Comprehensive surveys have presented different types of distributed MPC \cite{christofides2013distributed,negenborn2014distributed}, where one common approach reviewed in those surveys uses decomposition techniques to solve the MPC problem in a distributed fashion. Multiple decomposition techniques have been reported over the past decades based on Lagrangian \cite{geoffrion1974lagrangean,conejo2002decomposition}, Augmented Lagrangian \cite{kim1997coarse,cohen1980auxiliary,kim2000comparison} and Benders Decomposition \cite{shahidehopour2005benders}, which are generally based on the principle of decomposing the optimization problem into subproblems that can be solved in parallel. For the coordination of wind generation and storages, the Optimality Condition Decomposition (OCD) \cite{conejo2002decomposition,nogales2003decomposition} has been applied to solve the multi-step AC OPF problem successfully by dividing the entire power system into several regions each associated with a subproblem to solve \cite{Kyri2015MPC}. Other distributed MPC methods via dual decomposition \cite{wakasa2008decentralized} and temporal decomposition \cite{beccuti2004temporal} are also proposed. Distributed MPC not only reduces the computation burden in the centralized approach, but also helps preserve the data privacy among different control regions as only a small amount of data needs to be shared among neighboring regions to achieve the overall optimality of the entire system. 
 
However, while distributed approaches can alleviate the centralized computational burden, most distributed methods are iterative and generally take many iterations to converge, which may still lead to the violation of the time available for solving a specific problem. It has been observed that the number of iterations when using decomposition methods is greatly dependent on the system partitioning; i.e., which bus is assigned to which subsystem, if the overall system is decomposed geographically \cite{junyao2015impact}. Based on this observation, a partitioning method based on spectral clustering has recently been proposed that determines the best partition of a system such that the decomposition method can converge in fewer iterations using the determined partition \cite{guointelligent}. 
For improving the efficiency of the MPC approach, in this paper, we apply this partitioning method in conjunction with a decomposition technique to solve the MPC problem. Specifically, we consider optimizing the usage of the wind generation by using storage and employing a multi-step AC OPF problem that minimizes the total generation cost over a certain time horizon by optimally setting the charging/discharging status of the storages. Apart from the application considered in this paper, the proposed approach can also be applied to solve similar MPC-based multi-step optimization problems in a distributed fashion as well. The proposed distributed MPC consists of two steps: First, the partitioning method is applied to find the best partition of the test system; then, the multi-step optimization problem is solved using a decomposition method for a 24-hour time period. Through case studies, the time efficiency of the proposed distributed MPC approach is quantified and the impact of system partitioning on the speed of distributed MPC is highlighted. In particular, we demonstrate that the same partition can be used for solving the MPC problem for multiple time steps, which eases the practical use of distributed MPC as the partition of the system does not need to be changed frequently. 

The rest of the paper is organized as follows: In Section \ref{formulation}, the multi-step AC OPF problem is formulated and a distributed optimization problem formulation is also given. In Section \ref{methods}, the partitioning method and the decomposition method used in this paper are presented. Section \ref{casestudy} quantifies the effectiveness and efficiency of the distributed MPC approach with a focus on the impact of system partitioning on the convergence speed through simulations using the IEEE 14-bus and 118-bus test systems. Finally, Section \ref{conclusion} concludes the paper and proposes possible future directions.
\par}

\section{Problem Formulation}
\label{formulation}

\begin{figure}[htbp]
\setlength{\abovecaptionskip}{0.1cm} 
\centering
\includegraphics[trim = 18mm 0mm 18mm 0mm, clip=true,width=8cm, height=5cm]{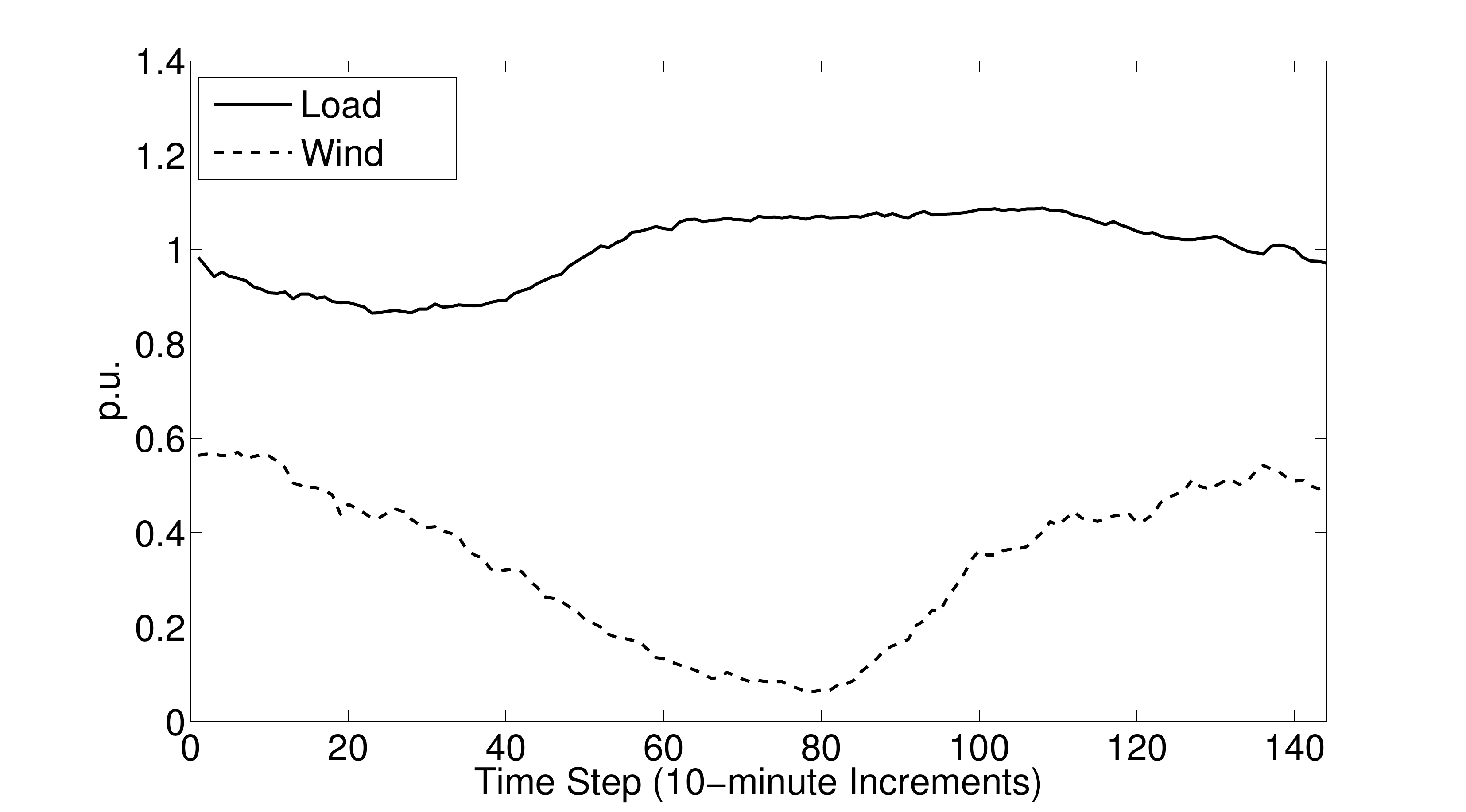} 
\caption{24-hour load and wind data with 10-minute intervals.}
\label{fig:loadwind}
\end{figure}
In this section, the centralized multi-step AC OPF problem is first formulated including wind generation and storages, and its formulation in the distributed form is then given. 
Figure \ref{fig:loadwind} shows an example of the wind and load data for a 24-hour period obtained from the Bonneville Power Administration at a 10-minute scale. As seen from Fig. \ref{fig:loadwind}, the wind generation at different times of the day is usually random and it is possible that the wind generation is high while the demand of the system is low. Hence, to reduce the generation cost at the peak demand, the procedure is to store the excessive energy generated by the cheap generators during the time when the wind generation can serve most of the system load and use the stored energy when the demand is high. With this purpose, we formulate a multi-step AC OPF problem where the objective is to minimize the total generation cost of non-renewable generations over a finite time horizon that consists of multiple time steps. The OPF problems at each time step are coupled by the charging and discharging of the storage devices. It is assumed that the wind generation must be consumed when available, hence it can be treated as negative load at the bus where the wind generation is placed. The overall multi-step AC OPF problem at time step $TS$ is formulated as follows:
\begin{subequations}
\begin{align}
\label{eqobjective}
\text{minimize}~~&f({P}_{G})=\sum_{t=TS}^{TS+N-1}\left(\sum_{i=1}^{G} \left(a_{i}P_{G_{i}}^{2}(t)+b_{i}P_{G_{i}}(t)+c_{i}\right)\right)\\
\nonumber\text{subject to}~~&\\
\label{eqactivepower}
&\nonumber\sum\limits_{i\in {{\Lambda }_{j}}}{{{P}_{{{G}_{i}}}}}(t)+P_{W_{j}}(t)-P_{In_{j}}(t)+P_{Out_{j}}(t)-{{P}_{{{D}_{j}}}}(t)=\\
&V_{j}(t)\sum\limits_{k\in {{\Omega }_{j}}}V_{k}(t)(g_{jk}\cos{{\theta }_{jk}}(t)+b_{jk}\sin{{\theta }_{jk}}(t))\\
\label{eqreactivepower}
&\nonumber \sum\limits_{i\in {{\Lambda }_{j}}}{{{Q}_{{{G}_{i}}}}}(t)-{{Q}_{{{D}_{j}}}}(t)=\\
&V_{j}(t)\sum\limits_{k\in {{\Omega }_{j}}}V_{k}(t)(g_{jk}\sin{{\theta }_{jk}}(t)-b_{jk}\cos{{\theta }_{jk}}(t))\\
\label{eqPV}
&E_{j}(t+T)=E_{j}(t)+\eta_c TP_{In_{j}}(t)-\frac{T}{\eta_d}P_{Out_{j}}(t)-\epsilon_{sbl}\\
\label{eqVlimit}
&V_{j}^{\min}\leq V_{j}(t)\leq V_{j}^{\max}\\
\label{eqPlimit}
&P_{G_{i}}^{\min}\leq P_{G_{i}}(t)\leq P_{G_{i}}^{\max}\\
&0\leq P_{In_{j}}(t)\leq P_{In_{j}}^{\max}\\
&0\leq P_{Out_{j}}(t)\leq P_{Out_{j}}^{\max}\\
&E^{\min}\leq E_{j}(t+T)\leq E^{\max}\\
\label{Ilimit}
&|I_{jk}|^{2}\leq (I_{jk}^{\max})^{2}
\end{align}
\end{subequations}
for $t = \{TS, \ldots, TS+N-1\}$ and $j = \{1, \ldots, B\}$.
The notations used in the problem formulation are listed below.

\begin{tabular}[htbp]{ll}
\label{tablenotation}
$N$ & optimization horizon \\
$B$ & number of buses\\
$G$ & number of generators\\
$a_i,b_i,c_i$ & cost parameters of generator $i$ \\
$P_{G_i}, Q_{G_i}$ & active and reactive power output of generator $i$ \\
$P_{D_{j}},Q_{D_{j}}$&active and reactive load at bus $j$\\
$P_{W_j}$ & active power output of wind generator at bus $i$\\
$P_{In_j}$ & power injected into storage at bus $j$\\
$P_{Out_j}$ & power drawn from storage at bus $j$\\
$V_{j}$&voltage magnitude of bus $j$\\
$\theta_{jk}$&difference of voltage angles between bus $j$ and bus $k$\\
$\Omega_j$ & set of buses connected to bus $j$ \\
$\Lambda_j$ & set of generators connected to bus $j$ \\
$E_j$ & energy level in the storage at bus $j$\\
$T$ & time between two consecutive time steps \\
$\eta$ &charging/discharging efficiency of storage at bus $j$\\
$\epsilon_{sbl}$&standby loss of storage\\
$\eta_c,\eta_d$&charging/discharging efficiency of the storage\\
$I_{jk}$&current on line from bus $j$ to bus $k$\\
\end{tabular}

\vspace{0.5cm}
Equations (\ref{eqactivepower}) and (\ref{eqreactivepower}) are the active and reactive power flow balances at each bus.  Equation (\ref{eqPV}) corresponds to the inter-temporal constraints on storages, (\ref{Ilimit}) reflects the line thermal limits and all other constraints denote the upper and lower limits on the variables. Apart from the constraints explicitly given above, the voltage angle at the slack bus is set to zero and the voltage magnitudes at generator buses are set to pre-determined values.
As a standard procedure in MPC, the solution found for the first time step is applied once the overall problem is solved. Then the optimization time horizon is shifted by time $T$ and the optimization problem is formulated and solved for the next time step. 

In the following, we formulate the problem (\ref{eqobjective}) to (\ref{Ilimit}) in a distributed fashion by grouping the variables into sets that correspond to different subproblems. Such a formulation will facilitate the implementation of decomposition methods, which will be explained in Section \ref{decomposition}. Note that the geographical decomposition of the problem is considered in this paper where there is a subproblem associated with each area. The reformulated optimization problem as a function of these sets of variables for a total of $K$ areas is given by
\begin{subequations}
\begin{align}
\label{eq9}
\underset{\bm{x}_{k}}{\text{minimize}}~~&\sum_{k=1}^{K} f_{k}(\bm{x}_{k})\\
\label{eq10}
\text{subject to}~~&c_{k}(\bm{x}_{1}, \dots,\bm{x}_{K})\leq 0;~ k=1,\dots,K\\
\label{eq11}
&n_{k}(\bm{x}_{k})\leq 0; ~k=1,\dots,K,
\end{align}
\end{subequations}
where $\bm{x_k}$ includes the variables assigned to subproblem $k$ and $f_k$ denotes the objective function associated with the $k$-th subproblem. Constraints (\ref{eqactivepower}) to (\ref{Ilimit}) are represented in a compact form by constraints (\ref{eq10}) and (\ref{eq11}). Constraint (\ref{eq10}) denotes the coupling constraint as it contains variables from multiple subproblems and (\ref{eq11}) denotes the non-coupling constraint as it only contains variables from one subproblem. In the considered OPF problem, the coupling constraints include the power flow balance at the buses placed at the boundaries of the areas and the thermal limits on tie lines connecting different areas, while all other constraints are considered as non-coupling constraints. The inequality constraints in (\ref{eq10}) and (\ref{eq11}) are handled with an Interior Point method.
  
\section{Partitioning and Decomposition}
\label{methods}
In this section, the two major methods used in our distributed optimization framework are introduced: 1) the partitioning method that determines the best geographical partition of the power system and 2) the decomposition method, namely, the Optimality Condition Decomposition method with Correction terms (OCD-C) that solves the OPF problem in a distributed fashion based on the partition determined using 1).  

\subsection{Decomposition Method}
\label{decomposition}
\subsubsection{Optimality Condition Decomposition}
\label{OCD}
Before introducing OCD-C, the general OCD method\cite{conejo2002decomposition,nogales2003decomposition} is presented. To solve (\ref{eq9})-(\ref{eq11}), one general approach is to derive the Lagrangian function which is denoted as $L$ and find the solutions to satisfy the KKT conditions. Denoting all the variables that need to be determined (including the Lagrange multipliers) by $\bm{y}$ and solving for the KKT conditions using the Newton-Raphson approach, the aforementioned procedure is equivalent to solving the following equations to get the update of variables $\Delta \bm{y}$ at each iteration:

\begin{equation}
\label{eq23}
\bm{H}\left( \begin{array}{c}
   \Delta \bm{y}_{1}  \\
   \vdots   \\
   \Delta \bm{y}_{K}  \\
\end{array} \right)=-\left( \begin{array}{c}
   \bm{KKT}_{1}  \\
   \vdots   \\
   \bm{KKT}_{N}\\
\end{array} \right)
\end{equation}
where
\begin{equation}
\bm{KKT_{k}}={\nabla }_{\bm{y}_{k}}L; k=1,...,K
\end{equation}
\begin{equation}
\label{eq24}
\bm{H}=\left( \begin{array}{ccc}
   \bm{H_{11}}&\dots &  \bm{H_{1K}}\\
   \vdots &\ddots&\vdots  \\
   \bm{H_{K1}}&\dots &  \bm{H_{KK}}\\
\end{array} \right)
\end{equation}
\begin{equation}
\label{eq25}
   \bm{H_{km}}=\nabla^{2}L_{\bm{y}_{k},\bm{y}_{m}};k,m=1,...,K.
\end{equation}
Here, the variables $\bm{y}$ are grouped according to subproblems and the indices in $\bm{y}_{1},...,\bm{y}_{K}$ indicate to which subproblem the variables belong. \(\bm{H}\) is the Hessian matrix of the Lagrangian function of the overall optimization function with the variables rearranged according to the subproblems they are assigned to. All the elements in the right-hand-side vector of (\ref{eq23}) have to be equal to zero at optimality. Notice that the subproblems are coupled by the non-trivial off-diagonal blocks $\bm{H_{km}}$ (where \(k\neq m\)) in \(\bm{H}\), which does not allow independent solutions of $\Delta \bm{y}_{1}$ to $\Delta \bm{y}_{K}$. Hence, to decouple the subproblems, OCD takes an approximate Newton step by setting the off-diagonal block elements in \(\bm{H}\) to zeros \cite{nogales2003decomposition} and then solving (\ref{eq23}). Consequently, each area can carry out the following Newton-Raphson step independently
\begin{equation}
\label{OCDupdate}
\Delta \bm{y}_{k}=-\bm{H_{kk}}^{-1}\cdot \bm{KKT}_{k}
\end{equation}

\noindent and update its variables \({\bm{y}}_{k}\leftarrow \bm{y}_{k}+\Delta \bm{y}_{k}\). Then the updated values of variables are shared among subproblems to enable the next iteration of calculation. Note that only a small number of the updated variables need to be exchanged between neighboring areas including the voltages of the buses placed at the boundaries, the Lagrange multipliers associated with the power flow balances at those buses and the multipliers associated with the tie line thermal limits.

\subsubsection{OCD with Correction Terms}
\label{OCDC}
Due to the fact that the updates of variables in OCD neglect the coupling between subproblems, it deviates from the centralized approach which results in more iterations until convergence. To alleviate this problem, an extended OCD with additional correction terms, namely, OCD-C is proposed and applied to various case studies \cite{junyao2015impact,Kyri2015MPC,guointelligent}. In OCD-C, the correction terms is added to the right-hand-side of (\ref{OCDupdate}) which results in the following updates of variables:
\begin{equation}
\label{OCDCupdate}
\Delta \bm{y}_{k}=\bm{H_{kk}}^{-1} \cdot (-\bm{KKT}_{k}+\hat{\bm{r}}_k)
\end{equation}
Here, $\hat{\bm{r}}_k$ is the correction term and it can be calculated by
\begin{equation}
\label{eq:ra}
\hat{\bm{r}}_k=\sum_{m=1,m\neq k}^{N} \bm{H_{km}} \bm{H_{mm}}^{-1} \cdot \bm{KKT}_{m}
\end{equation}
where each term in the summation can be calculated in one subproblem $m$ and sent to subproblem $k$. The detailed derivation of (\ref{eq:ra}) and the convergence criterion of OCD-C can be found in \cite{guointelligent} which shows that OCD-C can converge to the same solution as the centralized approach if the convergence criterion is fulfilled. As $\bm{H_{km}}$ and $\bm{r}_{k}$ are both sparse, the correction term only contains few non-zero terms which results in little additional information to be exchanged. However, by adding the correction term, OCD-C generally converges in notably fewer iterations than OCD and is used in this paper instead of OCD in the simulations.

\subsection{Power System Partitioning}
\label{IP}
It has been observed that the number of iterations until convergence when using OCD and OCD-C highly depends on how the system is partitioned into subsystems \cite{junyao2015impact}. Hence, to improve the time efficiency of decomposition methods, a partitioning method based on spectral clustering is proposed to determine a good partition of the system \cite{guointelligent}. In essence, the developed partitioning method defines a metric for measuring the computational coupling between buses based on the formulation of the considered optimization problem and groups the strongly computationally coupled buses into one subsystem. This is based on the premise that weaker couplings lead to less mutual impact among the subsystems, thus leading to faster convergence of the decomposition methods. It has been validated that the proposed partitioning method is effective in minimizing the computational coupling between subproblems to speed up the convergence of the decomposition methods. In the following, we present the rationale and key steps and procedures of the partitioning method, while more details can be found in \cite{guointelligent}.

As mentioned before, we first define an affinity metric between any two buses that measures or represents their computational coupling. The notation $\bm{H_{sys}}$ is used to denote the Hessian matrix of the overall Lagrangian function with variables ungrouped. We take advantage of the fact that if any entry \(H_{i,j}\) in $\bm{H_{sys}}$ is non-zero, this is an indication that the two variables with indices $i$ and $j$ are coupled; i.e., the updates of these two variables will appear in the same equation, hence, they directly affect each other. The larger the absolute value of \(H_{i,j}\), the stronger the coupling. Furthermore, it is assumed that the variables associated with one bus such as the voltage angle, the voltage magnitude and Lagrange multipliers should be assigned to the same subproblem. Hence, the affinity between any two buses is acquired based on the summation of all the absolute values of the elements in $\bm{H_{sys}}$ that are associated with these two buses. Again, a larger affinity denotes a stronger computational coupling. Specifically, the affinity metric $A_{m,n}$ between any bus $m$ and bus $n$ is calculated as follows:
\begin{equation}
\vspace{-0.1cm}
A_{m,n}=\underset{i\in S_{m}}\sum\underset{j\in S_{n}}\sum|H_{i,j}|+|Y_{m,n}|,
\label{eq:affinity}
\end{equation}
\noindent where $Y_{m,n}$ is the $(m,n)$-th element in the admittance matrix, and $S_{m}$ and $S_{n}$ denote the sets of the indices of the variables associated with buses $m$ and $n$, respectively. A more detailed explanation on the derivation of this affinity metric is given in \cite{guointelligent}.

After the affinity metric is calculated, the spectral clustering technique \cite{ng2002spectral} is applied which groups the buses based on the affinity among the buses. In this work, we pre-define the number of areas but then determine which bus should be assigned to which area. Note that: 1) the partition of the system only affects the assignment of variables into subproblems in the computation, but does not affect the \textit{physical partition of the power system}; 2) the partition of the system does not affect the exact solution of the optimization problem but only affects the time that decomposition methods take to converge to the solution; 3) the $\bm{H_{sys}}$ used in the calculation is evaluated at the optimal point which could be different depending on operating points. For the MPC problem, the operating point changes with the time step as the load and wind generation vary during a day. However, it will be shown in Section \ref{partitioning} that the partition of the system once determined is applicable to multiple time steps, hence does not need to be changed frequently. A simple explanation for this is that the affinity between buses is calculated mainly based on the line admittance, the voltage magnitude, the \(sin\) and \(cos\) of the differences between two bus angles, and the Lagrange multipliers for power flow and line thermal constraints, which do not change dramatically as the operating point changes if there is no severe line congestion. In Section \ref{partitioning}, it will be further discussed how to choose the operating point for applying the partitioning method in the MPC problem and handle the scenarios where there are different lines becoming congested.

\section{Case studies}
The distributed MPC approach is tested on the IEEE 14-bus and 118-bus systems. In this section, we will present two sets of simulation results. First, the results using different time horizons are compared, which show the benefit of the MPC approach in terms of reducing the generation cost and the ramping of the generators. Next, the convergence speeds of the decomposition method for different partitions are compared, which demonstrates the importance of system partitioning and the fact that the previously developed partitioning method can be effectively applied to the MPC problem.
\vspace{-0.1cm}
\label{casestudy}
\subsection{Simulation Setup}
\vspace{-0.1cm}
The simulations are run in Matlab on an iMac with 3.2GHz Intel Core i5 and 8GB memory.  The storage device has a roundtrip efficiency of $\eta_{c}\cdot\eta_{d}=0.95\%$, standby loss of 0.005 $p.u. \cdot $10-minute and maximum capacity of 1.0 $p.u. \cdot$ 10-minute. The wind and load data in Fig. \ref{fig:loadwind} is used. The simulations were run for a 24-hour period using the time horizons of $N=1,3,6$ and $9$ with the time interval of $T=10$ min, which correspond to no horizon, 30-minute, 60-minute, and 90-minute horizon, respectively. As the longest time horizon is 90 minutes and there are in total 144 time intervals over the 24-hour period, a total of 135 time steps are simulated using the available data. A multi-step AC OPF problem is solved at each time step. For comparison, the centralized optimization which uses the Newton-Raphson approach to update variables is also simulated. Convergence is achieved if the norm of all the mismatch between the constraints is lower than $10^{-3}$. The same starting point and convergence criterion are used for the OCD-C method with different partitions and the centralized approach. Note that the OCD-C always converges to the same solution as the centralized approach regardless of what partition is used. 
\vspace{-0.1cm}
\subsection{Impact of the Optimization Horizon}
\begin{figure}[htbp]
\setlength{\abovecaptionskip}{0.2cm} 
\centering
\includegraphics[trim = 0mm 0mm 0mm 0mm, clip=true,width=10cm]{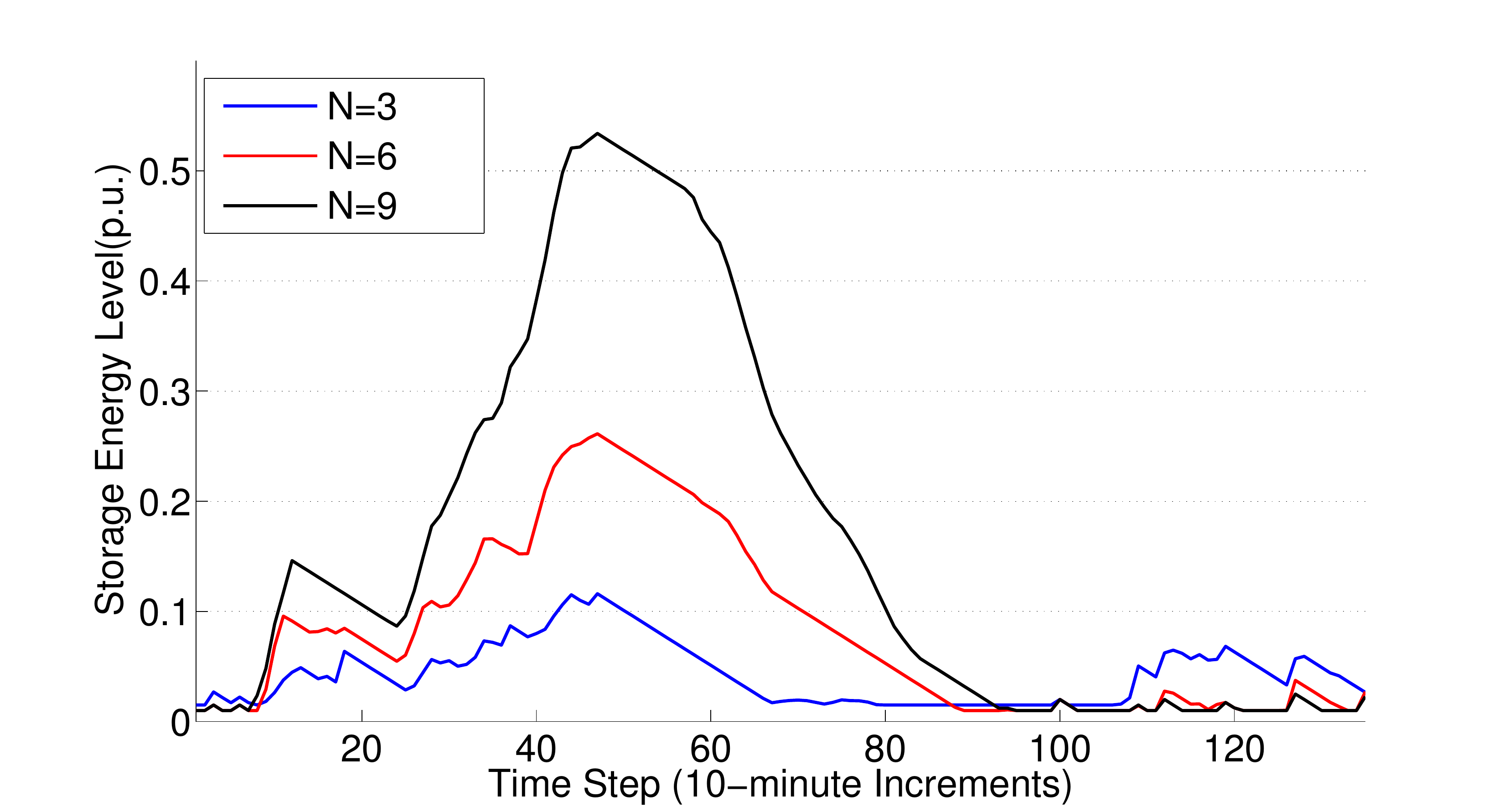} 
\caption{Optimal storage level of the storage device.}
\label{fig:storage}
\end{figure}

\vspace{-0.1cm}
In this subsection, the results for the IEEE 14-bus system are given for evaluating the impact of the length of the optimization horizon. A wind generator is located at Bus 5 and a storage device is located at Bus 14. Figure \ref{fig:storage} shows the optimal storage energy level with different time horizons denoted by $N$. It is clear that the utilization of the storage increases as the time horizon increases. The benefit of optimizing the usage of the storage is demonstrated in Table \ref{gencost} which shows the total generation cost and the total generator ramping over the simulated 24-hour period with different horizons. Again, as the time horizon increases, both the generator ramping and the generation cost decrease. Even though the generator ramping is not included as a hard constraint in the optimization problem, it has been reduced as the utilization of the storages smoothes out the fluctuations in the load. These results indicate that the MPC approach can effectively integrate the wind generation and storages especially with a longer time horizon. However, this does not indicate that one should extend the time horizon as much as possible, due to the fact that the problem size increases with the length of the time horizon, which will require more computation time and resources. Besides, the forecasted wind and load data might not be available or accurate for a long time horizon. Overall, the choice of the length of the time horizon depends on specific applications and the computation capability and is beyond the scope of this paper.

\begin{table}[htbp]
\caption{Total generator ramping and total generation cost}
\centering
\begin{tabular}{ p{5cm}<{\centering}| p{1cm}<{\centering}|p{1cm}<{\centering}| p{1cm}<{\centering}|p{1cm}<{\centering} }
\toprule
Time Horizon $N$&1&3&6&9\\
 \midrule
Total Generator Ramping (p.u.)&4.1604&3.9598&3.7841&3.7708\\
\midrule
Total Generation Cost (\$)&679,145&679,144&678,985&678,874\\
\bottomrule
\end{tabular}
\label{gencost}
\end{table}

\subsection{Impact of Partitioning}
\label{partitioning}
In this section, we focus on the efficiency of the distributed MPC approach and show that a good partition of the system is the key to efficiently implementing decomposition methods. To evaluate the performance of decomposition methods, two metrics are used, namely, the number of iterations \(n\) and the convergence time \(t\).  The convergence time is an approximation of the time spent on solving the subproblems in parallel. Specifically, \(t=n \cdot (\max \{t_{1} , t_{2}, ..., t_{K} \}) \) where \(t_{k}, k=1,...,K\) denotes the time spent on solving the $k$th subproblem at each iteration, which is assumed not to change much over iterations because the subproblem size stays the same. The time spent on information exchange is not accounted for in the current simulation, but will be investigated in future works. A smaller $n$ and $t$ denote a better partitioning of the system as the objective of the partitioning method is to reduce the iterations and computation time until convergence. 

\begin{figure}[htbp]
\setlength{\abovecaptionskip}{0.2cm} 
\centering
\captionsetup[subfigure]{captionskip=0cm}
\subfloat[SP Partition]
{
\label{fig:OP}
\includegraphics[trim = 12mm 0mm 14mm 0mm, clip=true,width=4.5cm]{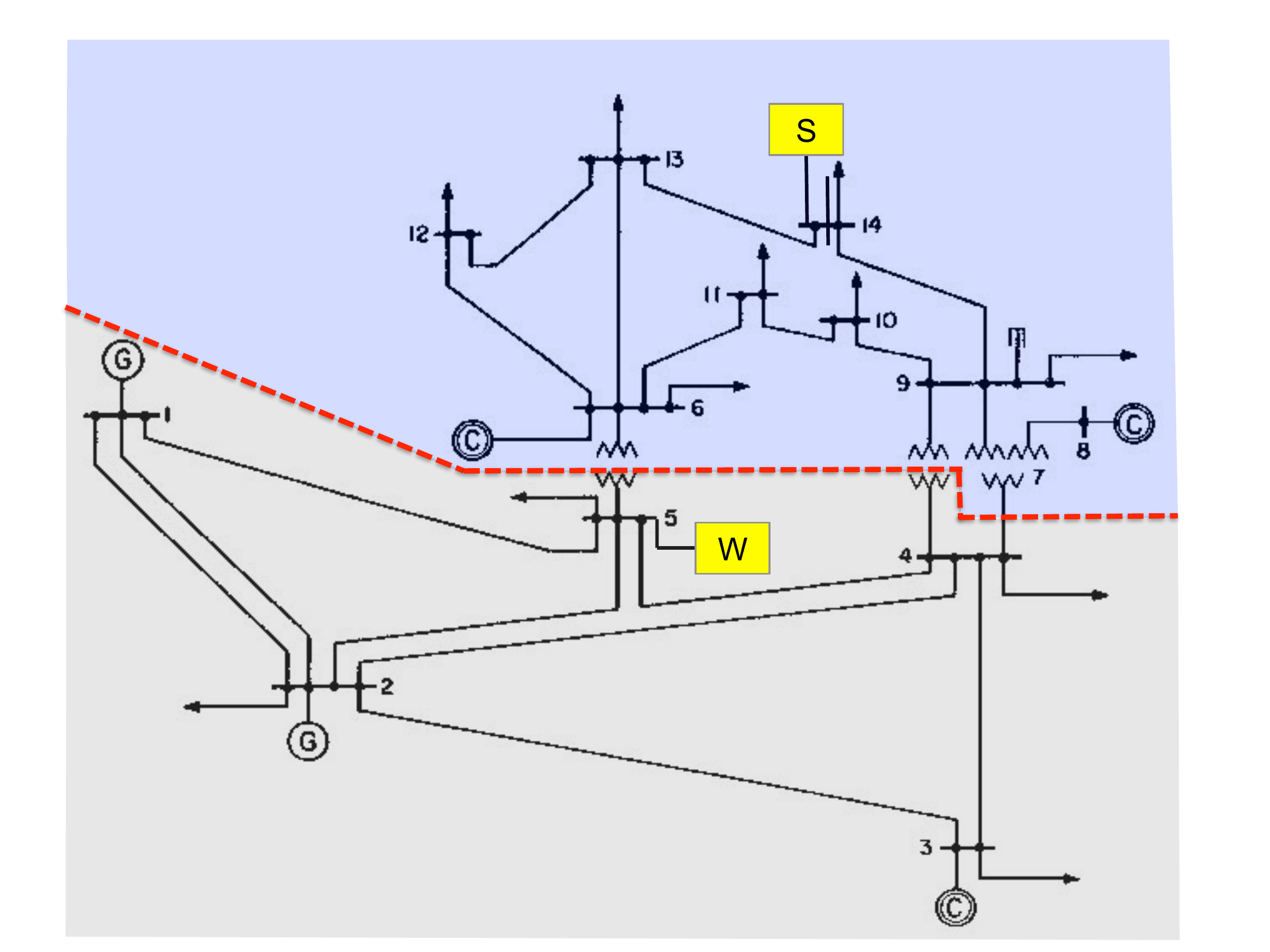} 
}
\hspace{0cm}
\subfloat[Arbitrary Partition]
{
\label{fig:NonOP}
\includegraphics[trim = 12mm 0mm 14mm 0mm, clip=true,width=4.5cm]{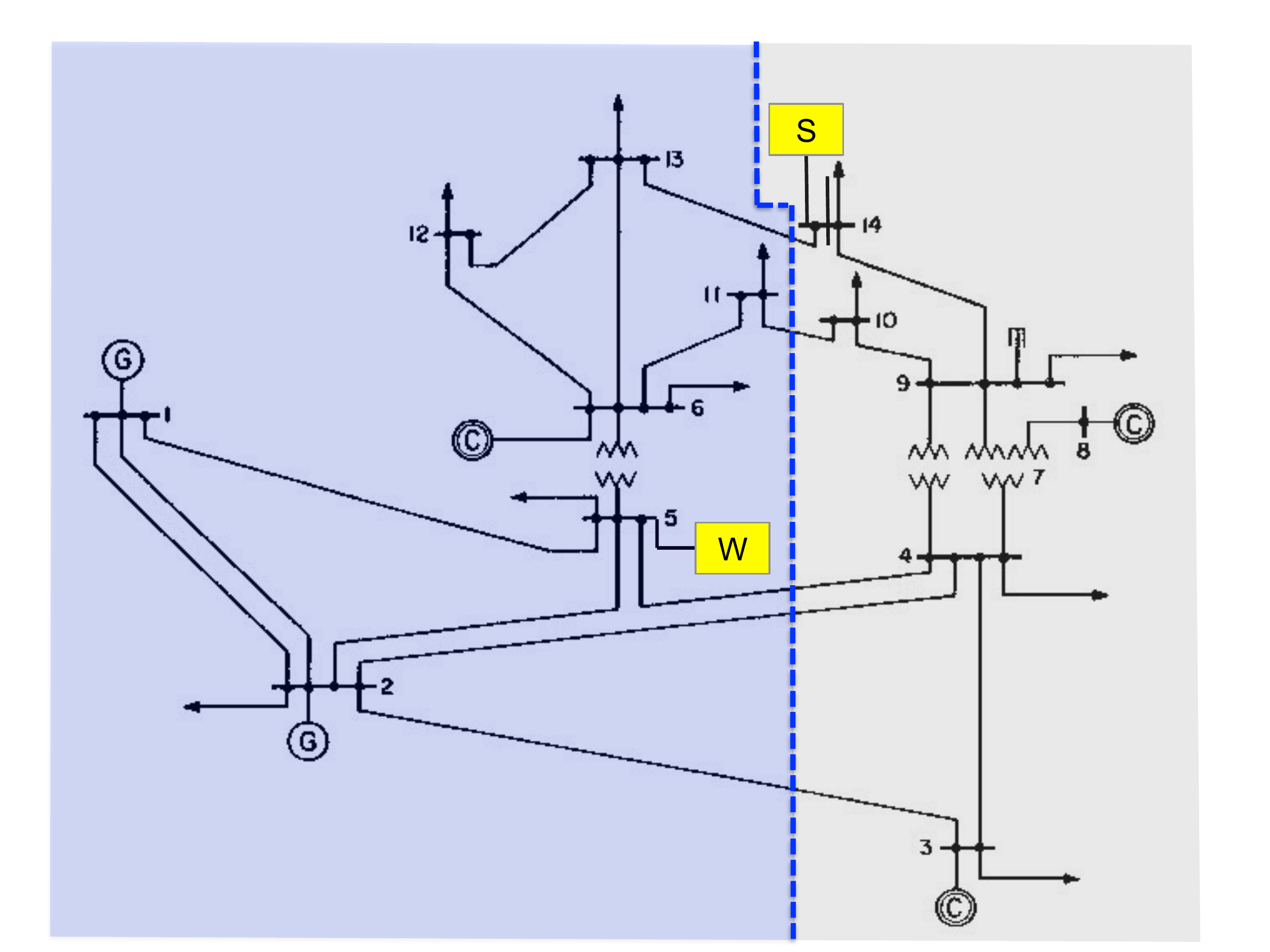} }
\caption{Two partitions of the IEEE-14 system.}
\label{fig:Partition14}
\end{figure}

For the 14-bus system, two partitions are used for the decomposition of the problem as shown in Fig. \ref{fig:Partition14}. ``SP Partition" denotes the best partition determined by the spectral partitioning technique presented in Section \ref{IP}, while ``Arbitrary Partition'' denotes an arbitrary geographical partition of the system. Note that it is highly likely that the arbitrary partition is chosen if one determines the partition only by observing the diagram of the system. The best partition is found at the operating point of base load level with $N=1$ and applied to solving the MPC problem at all time steps, which, as will be shown later, is also good for solving the MPC problem with an increased time horizon. 
\subsubsection{IEEE 14-bus system}

The average, median and maximum number of iterations until convergence of OCD-C using two different partitions are shown in Table \ref{fig:Iteration14} and compared with the centralized approach. As shown in Table \ref{fig:Iteration14}, using SP Partition will lead to significantly reduced iterations compared to the arbitrary partition. Note that decomposition methods would always take more iterations than the centralized method due to the fact that only partial information of the system is available at each subproblem and frequent information exchange needs to be made to achieve the overall optimality. 
\begin{table}[htb]
\caption{Number of iterations to convergence with the IEEE 14-bus system}
\centering
\begin{tabular}{ p{2cm}<{\centering}| p{0.5cm}<{\centering}|p{2cm}<{\centering}| p{2.5cm}<{\centering}|p{3cm}<{\centering} }
\toprule
Iterations& $N$&Centralized&SP Partition&Arbitrary Partition\\
 \midrule
Average&1&34&65&124\\
Median&1&34&73&135\\
Maximum&1&37&74&139\\
\midrule
Average&3&46&76&148\\
Median&3&46&83&156\\
Maximum&3&66&108&210\\
\midrule
Average&6&46&78&158\\
Median&6&46&84&159\\
Maximum&6&69&131&216\\
\midrule
Average&9&47&80&166\\
Median&9&46&85&166\\
Maximum&9&91&132&235\\
\bottomrule
\end{tabular}
\label{fig:Iteration14}
\end{table}

In terms of the actual computation time, the average convergence time is shown in Table \ref{fig:Time14}. It can be seen that using SP Partition, the convergence time of OCD-C is only slightly higher compared to the centralized approach, while the convergence using the arbitrary partition is much slower. We note here that for the IEEE-14 bus system, the centralized approach converges the fastest for most time steps due to the small problem size. However, when deployed on larger systems, the time efficiency of the distributed approach will be superior than that of the centralized approach, which will be demonstrated in Section \ref{118system}. Note that the sparsity of the matrices is exploited in the simulation to speed up the calculation, which works more to the advantage of the centralized approach where the matrices involved in calculations are relatively sparser. Hence, on this toy example, it is fairly impressive that the distributed MPC approach achieves a comparable time efficiency as the centralized approach if a good partition is used.
\begin{table}[b]
\caption{Average convergence time (in seconds) with the IEEE 14-bus system}
\centering
\begin{tabular}{ p{0.5cm}<{\centering}| p{2.5cm}<{\centering}|p{2.5cm}<{\centering}| p{3.5cm}<{\centering} }
\toprule
$N$&Centralized&SP Partition&Arbitrary Partition\\
 \midrule
1&0.015&0.027&0.040\\
3&0.072&0.093&0.164\\
6&0.211&0.223&0.380\\
9&0.348&0.408&0.695\\
\bottomrule
\end{tabular}
\label{fig:Time14}
\end{table}

\begin{figure}[htbp]
\setlength{\abovecaptionskip}{0cm} 
\centering
\includegraphics[trim = 0mm 0mm 0mm 15mm, clip=true,width=10cm]{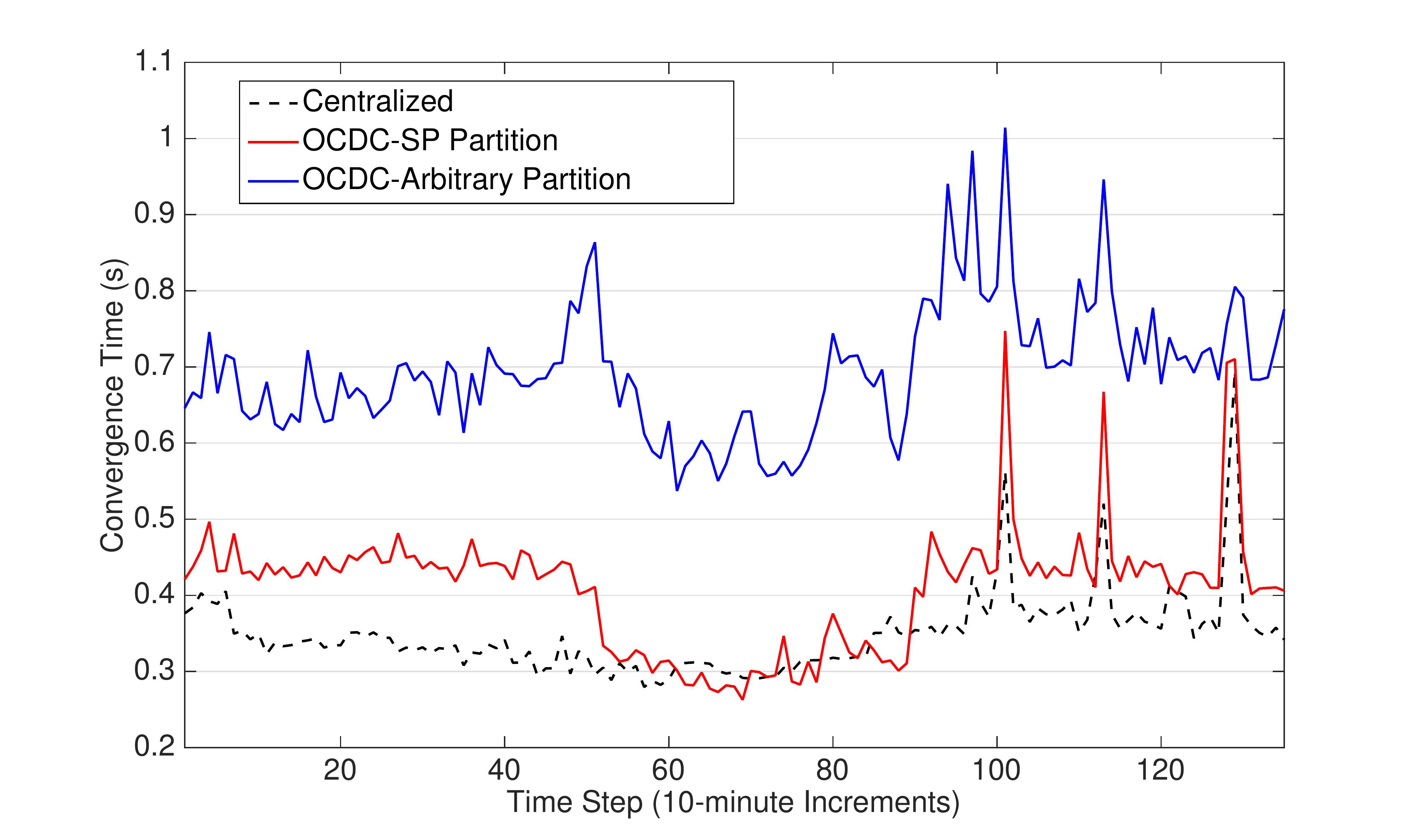} 
\caption{Convergence time with the IEEE 14-bus system.}
\label{fig:Convergence14}
\end{figure}

Now, to evaluate the robustness of the partition with respect to multiple time steps, the convergence time with the time horizon N=9 over all time steps are shown in Fig. \ref{fig:Convergence14}. In this 14-bus case, there is no line congestion observed over all time steps. As shown in Fig. \ref{fig:Convergence14}, SP Partition, which is the best partition found by the partitioning method at base load level for N=1, always leads to a reduced convergence time compared to Arbitrary Partition for all time steps. Hence, the best partition is fairly robust and there is no need to change the partition for different time steps in this case. 

\subsubsection{IEEE 118-bus system}
\label{118system}
\begin{figure}[t]
\setlength{\abovecaptionskip}{0.2cm} 
\centering
\captionsetup[subfigure]{captionskip=0cm}
\subfloat[SP Partition]
{
\label{fig:OP}
\includegraphics[trim = 20mm 0mm 20mm 0mm, clip=true,width=5cm]{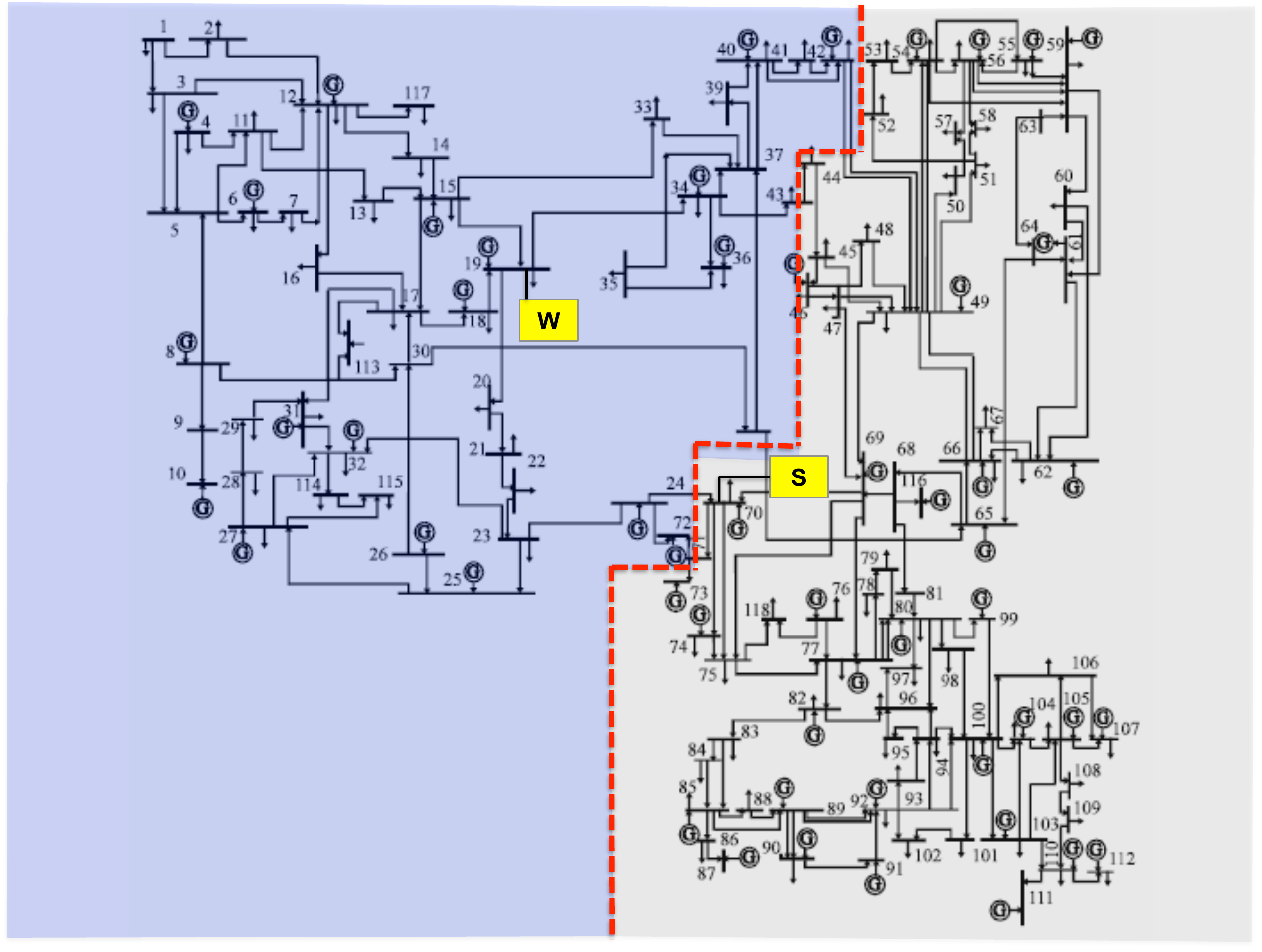} 
}
\hspace{0.5cm}
\subfloat[Arbitrary Partition]
{
\label{fig:NonOP}
\includegraphics[trim = 20mm 0mm 20mm 0mm, clip=true,width=5cm]{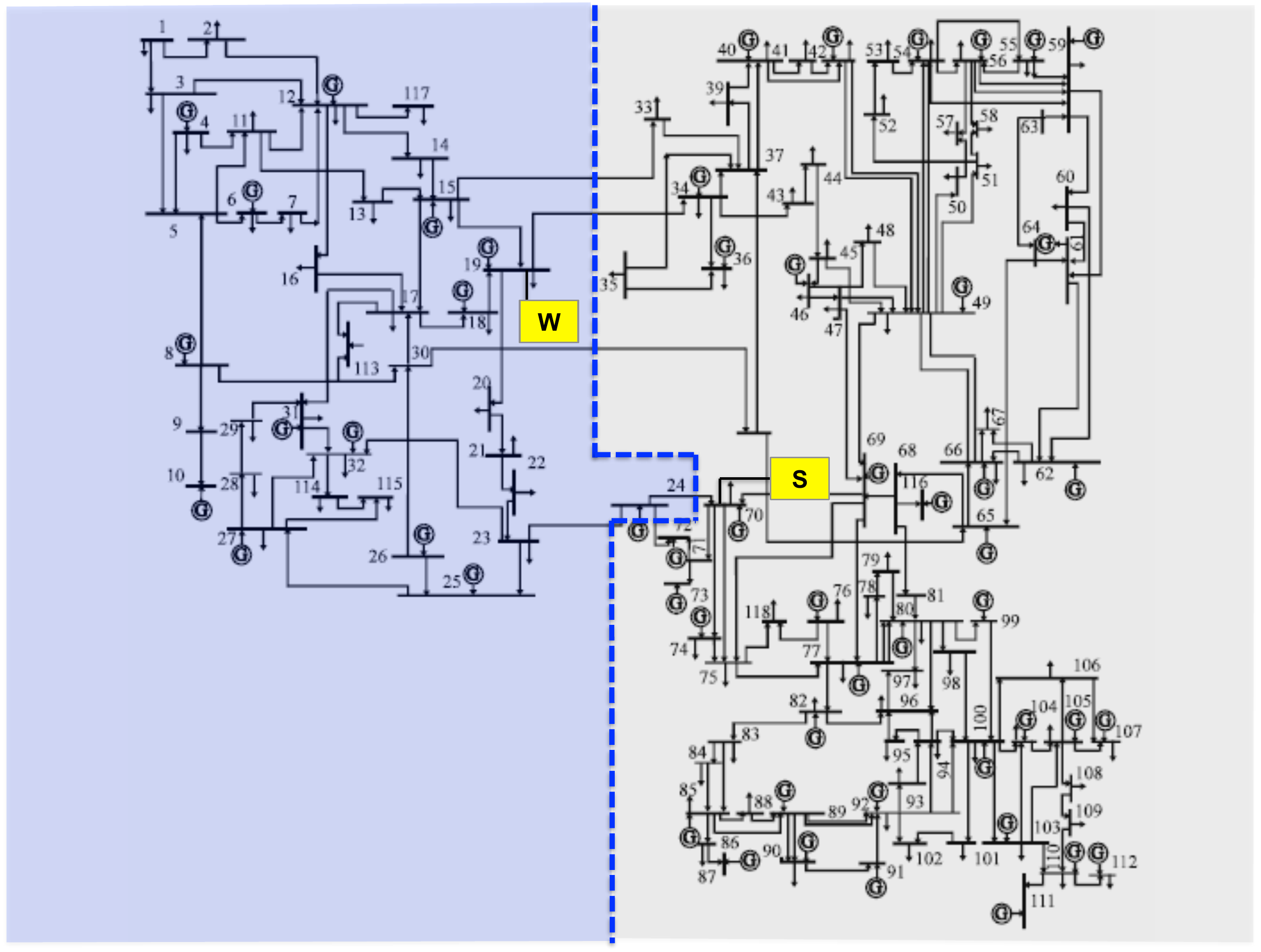} }
\caption{Two partitions of the IEEE-118 system.}
\label{fig:Partition118}
\end{figure}

To further test the efficiency of the distributed MPC approach, a larger system, namely the IEEE 118-bus system, is also used. A wind generator is located at Bus 19 and a storage device is located at Bus 70. Again, we use the partitioning method to find the best partition of the system and choose another arbitrary partition by observation of the system diagram for comparison. Both partitions are shown in Fig. \ref{fig:Partition118}. The number of iterations and time until convergence are shown in Table \ref{Iteration118} and \ref{Time118}, respectively. Similar to the 14-bus case, both the iterations and convergence time are significantly smaller using the best partition compared to the arbitrary one. It is worth highlighting that the average convergence time using the best partition is lower than the centralized approach, which shows an increased benefit of implementing distributed approaches on larger systems. 

\begin{table}[hbt]
\caption{Number of iterations to convergence with the IEEE 118-bus system}
\centering
\begin{tabular}{ p{2cm}<{\centering}| p{0.5cm}<{\centering}|p{2cm}<{\centering}| p{2.5cm}<{\centering}|p{3cm}<{\centering} }
\toprule
Iterations& $N$&Centralized&SP Partition&Arbitrary Partition\\
 \midrule
Average&1&43&61&102\\
Median&1&44&61&103\\
Maximum&1&57&64&127\\
\midrule
Average&3&54&72&128\\
Median&3&52&70&127\\
Maximum&3&83&97&176\\
\midrule
Average&6&54&76&145\\
Median&6&53&75&145\\
Maximum&6&84&106&225\\
\midrule
Average&9&56&80&157\\
Median&9&54&77&154\\
Maximum&9&81&112&264\\
\bottomrule
\end{tabular}
\label{Iteration118}
\end{table}
\begin{table}[hbt]
\caption{Average convergence time (in seconds) with the IEEE 118-bus system}
\centering
\begin{tabular}{ p{0.5cm}<{\centering}| p{2.5cm}<{\centering}|p{2.5cm}<{\centering}| p{3.5cm}<{\centering} }
\toprule
$N$&Centralized&SP Partition&Arbitrary Partition\\
 \midrule
1&0.287&0.256&0.528\\
3&2.210&1.766&3.861\\
6&7.689&7.112&16.297\\
9&17.247&16.173&38.254\\
\bottomrule
\end{tabular}
\label{Time118}
\end{table}

\begin{figure}[tb]
\setlength{\abovecaptionskip}{0cm} 
\centering
\includegraphics[trim = 0mm 0mm 0mm 0mm, clip=true,width=10cm]{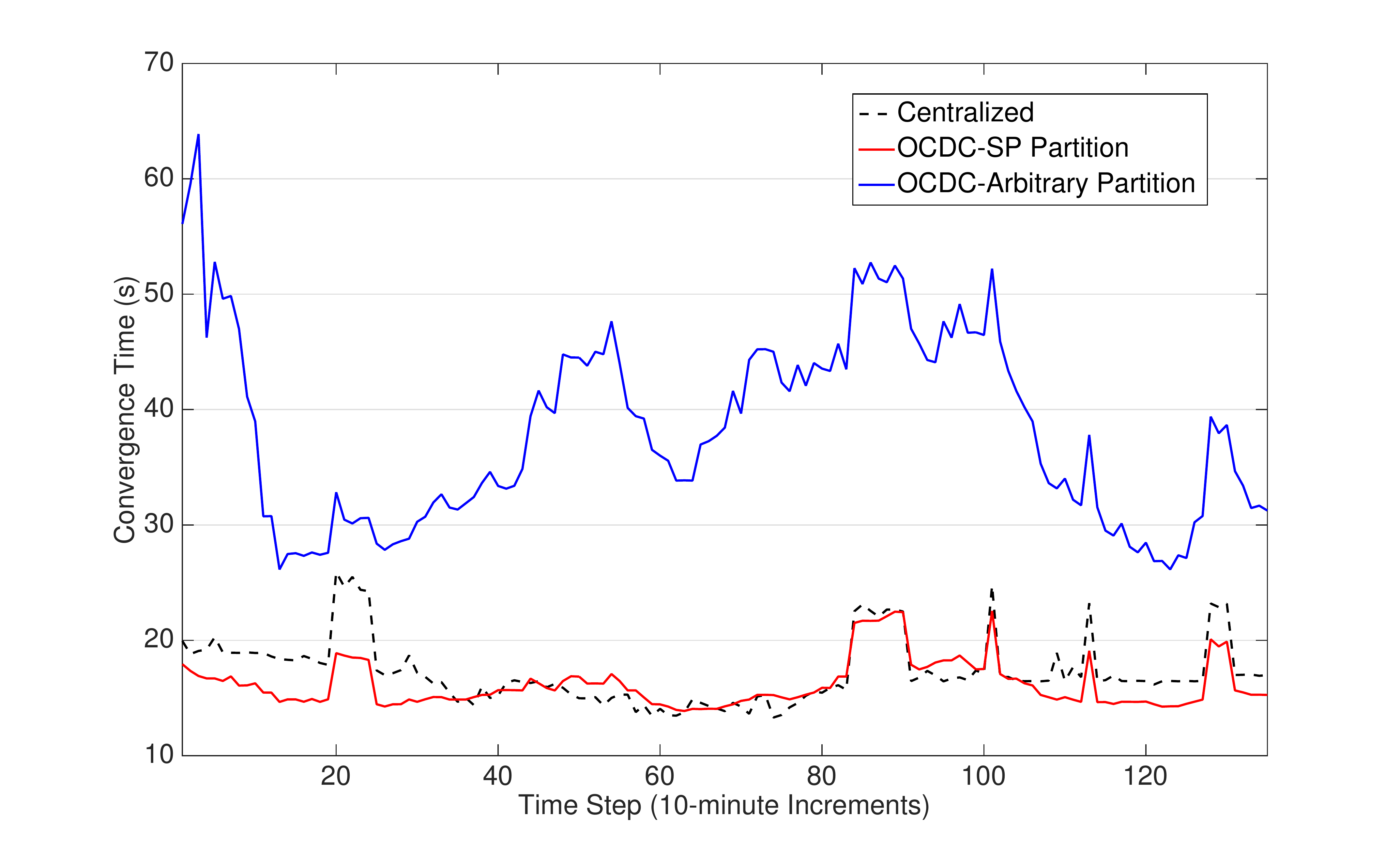} 
\caption{Convergence time with the IEEE 118-bus system.}
\label{fig:Convergence118}
\end{figure}

The convergence time with the time horizon N=9 over all time steps are shown in Fig. \ref{fig:Convergence118}. Note that for time step 1 to 13, a tie line associated with the best partition becomes congested, which, however, does not affect the convergence time of the OCD-C method much. In other words, the best partition is still robust with all time steps even when line congestion occurs. However, there could be cases where the convergence performance of the decomposition method is degraded once the tie line constraints become binding due to the increased computational coupling between the two areas that the tie line connects. 

Here, we provide insights into how one can determine the partition of a system for different time steps when the load levels are different. Since the best partition performs well for a fairly large range of load levels, one can find the best partition at the load level that occurs during most of the day. When the tie lines associated with the best partition are not severely congested as in the considered case, the best partition can be applied to all time steps. When the tie lines become severely congested, the partitioning method can be applied for that particular operating point to find a new partition. Overall, due to the robustness of the best partition, it can be expected that the computation effort spent on determining the partition of the system is quite low because the partition only needs to be computed for several operating points with different line congestion scenarios. 
%

%
\section{Conclusions}
\label{conclusion}
In this paper, we applied a partitioning method in conjunction with a decomposition technique to solve a MPC-based multi-step AC OPF problem in a distributed manner, which results in an effective integration of the wind generation and the storage device. Through simulation results, we showed that by determining a good partition of the system using the presented partitioning method, the efficiency of the decomposition method can be significantly improved. In particular, the computation time using the proposed distributed approach is shorter compared with the centralized approach when applied to large systems. Furthermore, the best partition of the system is applicable to a wide range of time steps. The proposed distributed optimization approach can also be used for solving other general multi-step optimization problems, which provides a useful tool in the planning and management of power systems.

For future work, we plan to investigate how the information exchange involved in the distributed optimization can be implemented in real systems and how the associated communications latency affects the overall efficiency of decomposition methods.

{\renewcommand\baselinestretch{1.3}\selectfont
\section*{Acknowledgment}
The authors would like to thank ABB for the financial support and particularly Dr. Xiaoming Feng for his invaluable inputs.

\bibliography{Partition}
\par}

\end{document}